\documentclass[aps,preprint]{revtex4}%
\usepackage{amsmath}
\usepackage{graphicx}
\usepackage{amsfonts}
\usepackage{amssymb}%
\begin{document}
\preprint{ }
\title{Equilibrium properties of self-interacting neutrinos in the quasi-particle approach}
\author{M. Sirera}
\affiliation{Departamento de Astronom\'{\i}a y Astrof\'{\i}sica, Universidad de Valencia,
46100 Burjassot (Valencia) Spain}
\author{A. P\'{e}rez}
\affiliation{Departamento de F\'{\i}sica Te\'{o}rica and IFIC Universidad de Valencia,
46100 Burjassot (Valencia) Spain}

\begin{abstract}
In this work a neutrino gas in equilibrium is studied both at $T=0$ and at
finite temperature. Neutrinos are treated as massive Dirac quasi-particles
with two generations. We include self-interactions among the neutrinos via
neutral currents, as well as the interaction with a background of matter. To
obtain the equilibrium properties we use Wigner function techniques. To
account for corrections beyond the Hartree approximation, we also introduce
correlation functions. We prove that, under the quasi-particle approximation,
these correlation functions can be expressed as products of Wigner functions.
We analyze the main properties of the neutrino eigenmodes in the medium, such
as effective masses and mixing angle. We show that the formulae describing
these quantities will differ with respect to the case with no self-interactions.

\end{abstract}
\maketitle

\section{Introduction.}

Propagation in dense media is one of the most interesting issues in present
neutrino physics. The consequences on solar and atmospheric neutrino data, as
well as in baseline neutrino experiments, are crucial for the understanding of
neutrino properties. Moreover, in astrophysical and cosmological scenarios,
they provide an important feedback which modifies the physical evolution of
the system under consideration. This, in fact, is the case for supernova
explosions, where neutrino interactions and oscillations can change the shock
dynamics \cite{1,2,3}.

The nucleosynthesis of heavy elements via r-processes is also largely affected
by neutrino oscillations, since the proton to neutron ratio can be altered by
the oscillations.

In a similar way, they have been shown to play an important role in
establishing the cosmic flavor content when the Big Bang nucleosynthesis
starts, thus influencing the helium production. A crucial ingredient in these
two scenarios is the neutrino self- interaction \cite{4,5,6,3l}. As it has
been remarked, such interactions are non-diagonal in flavor space, and give
rise to new phenomena in the oscillatory behavior \cite{7,8,9}.

In this work, we analyze in detail the \emph{equilibrium} properties of a
self-interacting neutrino gas. Here, we study a system consisting on two
generations of neutrinos in equilibrium with a matter background and
self-interactions taken into account. The formalism, however, can be extended
in a straightforward way to the case of three generations.

We assume that chemical equilibrium has been reached, therefore the chemical
potential is the same for both neutrino species. This is, in fact, the
situation for muon and tau neutrinos inside a supernova, where both kinds of
neutrinos are produced in pairs and, therefore, the chemical potentials are
zero. It seems also the case in the early Universe, with the present values of
mass differences and mixing angles, where chemical equilibrium is achieved (at
least approximately) just before the nucleosynthesis epoch for the three
neutrino families \cite{4}.

Neutrinos will be treated in the \emph{quasi-particle approximation}, i.e. we
assume that the field corresponding to the mass eigenstates can be described
as a superposition of plane waves with a modified (with respect to the vacuum)
dispersion relation. As we will show, the equilibrium features, such as the
effective masses and mixing angles, are modified when the interactions among
neutrinos are considered, due to the non-diagonal nature of the self-term. To
this end, we use a method based on \emph{Wigner Functions}, which has been
shown to be appropriate to describe both the equilibrium and kinematics of
many-particle neutrino mixing \cite{10}. To account for corrections beyond the
Hartree approximation, we also introduce correlation functions. We prove that,
under the quasi-particle approximation, these correlation functions can be
expressed as products of Wigner functions. As we show, the results obtained
using these techniques agree with previous calculations \cite{SR93,R96,3p,3s}.

This paper is organized as follows. In section II, we derive the equations
satisfied by the Wigner functions of self-interacting massive neutrinos. In
section III we consider the Hartree approximation (i.e., when correlations are
neglected). In this case, the interactions among neutrinos contribute in the
same way as the neutral current interactions of neutrinos with other
particles, such as electrons, protons or neutrons. The addition of
correlations results in the appearance of non-lineal effects due to the
self-interaction. This is described in section IV. In section V we analyze the
resulting dispersion relations of eigenmodes. Our main results are summarized
in section VI. The appendix at the end contains the derivation of the
statistical correlations used in section IV.

\section{Equations for Wigner functions.}

We consider a neutrino gas consisting on two generations of neutrinos which
interact with themselves by means of neutral current interactions. We denote
the two flavors by (e,$\mu$). Of course, the formalism can also be applied to
any two flavors, such as $\mu$ and $\tau$ neutrinos.

Since we deal with two neutrino species, it is convenient to introduce vectors
and matrices in flavor (or mass) space. We therefore define the neutrino and
antineutrino vector fields:
\begin{equation}
\widehat{\nu}(x)\equiv\left(
\begin{array}
[c]{c}%
\widehat{\nu}^{e}(x)\\
\widehat{\nu}^{\mu}(x)
\end{array}
\right)  ,\ \ \widehat{\overline{\nu}}(x)\equiv\left(
\begin{array}
[c]{cc}%
\widehat{\overline{\nu}}^{e}(x) & \widehat{\overline{\nu}}^{\mu}(x)
\end{array}
\right)  \label{21}%
\end{equation}

The symbol $\ \symbol{94}$ on top of a magnitude means that we are dealing
with a quantum operator. This will be used to distinguish this magnitudes from
statistical averages. We also introduce the following matrices in flavor space
\begin{align}
\Lambda^{ab,\mu} &  \equiv\left(
\begin{array}
[c]{cc}%
0 & 0\\
0 & \gamma^{\mu}(1-\gamma^{5})
\end{array}
\right)  \label{24}\\
\quad\Omega^{ab,\mu} &  \equiv\left(
\begin{array}
[c]{cc}%
\gamma^{\mu}(1-\gamma^{5}) & 0\\
0 & 0
\end{array}
\right)
\end{align}

With the aid of these notations, the Lagrangian terms which account for the
neutrino self-interactions \cite{7} can be written as
\begin{align}
\widehat{\mathit{L}}(x) &  =\widehat{\overline{\nu}}(x)i\gamma^{\mu}%
\partial_{\mu}\widehat{\nu}(x)-\widehat{\overline{\nu}}(x)M\widehat{\nu
}(x)-\frac{G_{F}}{4\sqrt{2}}\widehat{\overline{\nu}}(x)\Omega^{\mu}%
\widehat{\nu}(x)\widehat{\overline{\nu}}(x)\Omega_{\mu}\widehat{\nu
}(x)\nonumber\\
- &  \frac{G_{F}}{4\sqrt{2}}\widehat{\overline{\nu}}(x)\Lambda^{\mu}%
\widehat{\nu}(x)\widehat{\overline{\nu}}(x)\Lambda_{\mu}\widehat{\nu}%
(x)-\frac{G_{F}}{2\sqrt{2}}\widehat{\overline{\nu}}(x)\Omega^{\mu}\widehat
{\nu}(x)\widehat{\overline{\nu}}(x)\Lambda_{\mu}\widehat{\nu}(x),
\end{align}
where $G_{F}$ is the Fermi constant and $M$ is the neutrino mass matrix. The
equation of motion then reads as%

\begin{align}
&  i\gamma^{\mu}\partial_{\mu}\widehat{\nu}(x)-M\widehat{\nu}(x)-\frac{G_{F}%
}{4\sqrt{2}}\Omega^{\mu}\widehat{\nu}(x)\widehat{\overline{\nu}}(x)\Omega
_{\mu}\widehat{\nu}(x)\nonumber\\
-  &  \frac{G_{F}}{4\sqrt{2}}\widehat{\overline{\nu}}(x)\Omega^{\mu}%
\widehat{\nu}(x)\Omega_{\mu}\widehat{\nu}(x)-\frac{G_{F}}{4\sqrt{2}}%
\Lambda^{\mu}\widehat{\nu}(x)\widehat{\overline{\nu}}(x)\Lambda_{\mu}%
\widehat{\nu}(x)\nonumber\\
&  -\frac{G_{F}}{4\sqrt{2}}\widehat{\overline{\nu}}(x)\Lambda^{\mu}%
\widehat{\nu}(x)\Lambda_{\mu}\widehat{\nu}(x)-\frac{G_{F}}{2\sqrt{2}}%
\Omega^{\mu}\widehat{\nu}(x)\widehat{\overline{\nu}}(x)\Lambda_{\mu}%
\widehat{\nu}(x)\nonumber\\
-  &  \frac{G_{F}}{2\sqrt{2}}\widehat{\overline{\nu}}(x)\Omega^{\mu}%
\widehat{\nu}(x)\Lambda_{\mu}\widehat{\nu}(x)=0
\end{align}

(a similar equation can be derived for the adjoint field $\widehat
{\overline{\nu}}(x)$ ).

We now introduce the neutrino \emph{Wigner} \emph{operator}
\begin{equation}
\widehat{F}_{ij}^{ab}(x,p)=(2\pi)^{-4}\int d^{4}y\;e^{-ipy}\,\widehat
{\overline{\nu}}_{j}^{b}(x+y/2)\widehat{\nu}_{i}^{a}(x-y/2) \label{wigopneu}%
\end{equation}
From Eq. (\ref{wigopneu}) one can easily show that the Hermitian conjugate is
given by
\begin{equation}
\widehat{F}_{ij}^{ab\dagger}(x,p)=\gamma_{jk}^{0}\widehat{F}_{kq}%
^{ba}(x,p)\gamma_{qi}^{0} \label{216}%
\end{equation}

One obtains the following equation for the neutrino Wigner Operator :
\begin{align}
&  \gamma\lbrack\partial\widehat{F}(x,p)-2ip\widehat{F}(x,p)]+2iM\widehat
{F}(x,p)=-(2\pi)^{-4}\frac{iG_{F}}{2\sqrt{2}}\int d^{4}y^{\prime}%
d^{4}k\,e^{-ik(x-y^{\prime})}\nonumber\\
&  [\Omega\widehat{F}(x,p-k/2)\widehat{\overline{\nu}}(y^{\prime})\Psi
\widehat{\nu}(y^{\prime})+\widehat{\overline{\nu}}(y^{\prime})\Phi\widehat
{\nu}(y^{\prime})\Lambda\widehat{F}(x,p-k/2)\nonumber\\
&  +\widehat{\overline{\nu}}(y^{\prime})\Omega\widehat{\nu}(y^{\prime}%
)\Omega\widehat{F}(x,p-k/2)+\Lambda\widehat{F}(x,p-k/2)\widehat{\overline{\nu
}}(y^{\prime})\Lambda\widehat{\nu}(y^{\prime})] \label{eqwopneu}%
\end{align}
where we defined the matrices $\Psi=\Omega+2\Lambda$ and $\Phi=2\Omega
+\Lambda$.

We are now interested in introducing statistical averages from the quantum
operators defined above. These statistical averages are called \emph{Wigner
functions} \cite{10}, and are the analogous to the distribution functions we
need to describe many-particles systems. These are, in general, complex
functions, and also contain a Lorentz structure.
The neutrino Wigner functions are defined as:%
\begin{equation}
F_{ij}^{ab}(x,p)\equiv<\widehat{F}_{ij}^{ab}(x,p)>=(2\pi)^{-4}\int
d^{4}y\,e^{-ipy}<\widehat{\bar{\nu}}_{j}^{b}(x+y/2)\widehat{\bar{\nu}}_{i}%
^{a}(x-y/2)>
\end{equation}

Here, the symbol $<\widehat{A}>$ means the average of a given quantum operator
$\widehat{A}$ over a basis of quantum states which are compatible with the
macroscopical knowledge of the system. The latter determines a given density
matrix operator $\widehat{\rho}$ . Thus the averaging is performed according
to
\begin{equation}
<\widehat{A}>\equiv Sp\{\widehat{\rho}\widehat{A}\} \label{average}%
\end{equation}

where $Sp$ means the trace performed over the quantum basis. Taking into
account Eq. (\ref{216}) one immediately obtains :
\begin{equation}
F^{\dagger}(x,p)=\gamma^{0}F(x,p)\gamma^{0} \label{hermitianF}%
\end{equation}
which implies that $\widehat{F}(x,p)$ is an Hermitian matrix with respect to
generation indices.

Starting from Eq. (\ref{eqwopneu}), one can take statistical averages to
obtain the equations of motion for the Wigner Function, which turns out to be%

\begin{align}
&  [\gamma(\partial-2ip)+2iM]F(x,p)=-(2\pi)^{-4}\frac{iG_{F}}{\sqrt{2}}\int
d^{4}y^{\prime}d^{4}kd^{4}k^{\prime}\,e^{-ik(x-y^{\prime})}\nonumber\\
\lbrack &  <\Omega\widehat{F}(x,p-k/2)Tr\Psi\widehat{F}(y^{\prime},k^{\prime
})>+<Tr\Psi\widehat{F}(y^{\prime},k^{\prime})\Lambda\widehat{F}%
(x,p-k/2)>\nonumber\\
+ &  <Tr\Omega\widehat{F}(y^{\prime},k^{\prime})\Omega\widehat{F}%
(x,p-k/2)>+<\Lambda\widehat{F}(x,p-k/2)Tr\Lambda\widehat{F}(y^{\prime
},k^{\prime})>]\label{eqforWig}%
\end{align}

\section{Neutrino gas in the Hartree approximation}

In this section we investigate the neutrino gas under the assumption of global
equilibrium. This means that macroscopic quantities must be invariant under
space-time translations:
\begin{equation}
F(x,p)=F(p)
\end{equation}
Moreover, as a first approximation, we neglect the effect of statistical
correlations, which simply translates into replacing the statistical average
of an operator product by the product of their averages:
\begin{equation}
<\widehat{F}_{ij}^{ab}(p)\widehat{F}_{kl}^{cd}(p^{\prime})>=F_{ij}%
^{ab}(p)F_{kl}^{cd}(p^{\prime})
\end{equation}
Under these conditions, we obtain that the equation of motion for the Wigner
Function of the neutrinos in equilibrium is \cite{10}
\begin{equation}
(\gamma p-M)F(p)=\frac{G_{F}}{2\sqrt{2}}\int d^{4}kTr[\gamma_{\mu}%
(1-\gamma^{5})F(k)]\gamma^{\mu}(1-\gamma^{5})F(p),\label{HartreeEqui}%
\end{equation}

In order to obtain a solution to the problem, we simplify our approach to the
neutrino gas by making the following assumptions:

1) Neutrinos are assumed to have a small mass (as compared to typical
energies). Therefore, the neutrino fields consist essentially on
left-chirality projections, i.e. we assume that the right projections are very
small, as compared to the left projections. Moreover, we can consider that the
neutrino gas contains only neutrinos with negative polarization (negative
helicity) and antineutrinos with positive polarization (positive helicity).
Thus we have, to a good approximation, that
\begin{equation}
F(p)\simeq F_{L}(p)\simeq F^{-}(p)+\overline{F}^{+}(p). \label{PrimeHipo}%
\end{equation}

2) Quasi-particle approximation. In the equilibrium gas, neutrinos are in
their interacting eigenstates, i.e. a particle with definite momentum and
polarization has also a definite energy. This allows us to treat the neutrinos
of the ensemble as free particles with an \emph{effective mass} instead of
their own masses in vacuum. In other words, the field can be decomposed in
plane waves, with a dispersion relation which differs from the one in vacuum.

3) The mixing angle, which allows us to write a flavor eigenstate as a linear
combination of effective mass eigenstates, is supposed in principle to depend
on all the degrees of freedom of the particle, in a similar way to what
happens with the mixing angle in matter, when neutrinos are propagating trough
a matter background, but do not interact among them.

With all these hypothesis, we can now rewrite the equation of motion as
\begin{equation}
(\gamma p-M)F_{L}(p)=\sqrt{2}G_{F}\gamma^{\mu}\int d^{4}kTr\gamma_{\mu}%
F_{L}(k)F_{L}(p),\label{HartreeEqui1}%
\end{equation}
or, in a more simplified way,
\begin{equation}
(\gamma p-\sqrt{2}G_{F}\gamma a-M)F_{L}(p)=0,\label{HartreeEqui2}%
\end{equation}
where the four-vector $a$ is defined as
\begin{equation}
a_{\mu}=\int d^{4}kTr\gamma_{\mu}F_{L}(k).\label{constant a}%
\end{equation}

In order to obtain an equation which gives the \emph{propagation modes }of the
neutrinos in the gas, we multiply the above formula by $(\gamma p-\sqrt
{2}G_{F}\gamma a+M),$ and neglect the subdominant terms proportional to
$G_{F}^{2}$. We then arrive to the equation:
\begin{equation}
(p^{2}-2\sqrt{2}G_{F}ap-M^{2})F_{L}(p)=0.\label{propaequa}%
\end{equation}
Without any loss of generality, we can take $p^{\mu}=(p^{0},|\overrightarrow
{p}|,0,0)$ , therefore the interaction term is $ap=a^{0}p^{0}-a^{1}|\vec{p}|$.
On the other hand, using the known properties of the Wigner Functions
\cite{10} $\widetilde{F}_{L}^{-}$ and $\widetilde{\overline{F}}_{L}^{+}$
corresponding to mass eigenstates of neutrinos and antineutrinos,
respectively, we obtain
\begin{align}
Tr\gamma^{\mu}F_{L}(k) &  =Tr\gamma^{\mu}U(k)\widetilde{F}_{L}(k)U^{\dagger
}(k)=Tr\gamma^{\mu}\widetilde{F}_{L}(k)=\nonumber\\
&  =Tr\gamma^{\mu}\sum_{a=1,2}[\widetilde{F}_{L}^{-aa}(k)+\widetilde
{\overline{F}}_{L}^{+aa}(k)],\label{trace}%
\end{align}
where $U$ is the unitary transformation that transforms from mass eigenstates
to flavor states. After working out the traces and the integrations, one
obtains that the interaction term is $ap=n_{\nu}p^{0},$ where $n_{\nu}%
=n_{1}+n_{2}=n_{\widetilde{\nu}_{1}}+$ \ $n_{\widetilde{\nu}_{2}%
}-n_{\widetilde{\overline{\nu}}_{1}}-n_{\widetilde{\overline{\nu}}_{2}}$ is
the total density of neutrinos (with $p^{0}\simeq|\overrightarrow{p}|$ ) minus
the corresponding antineutrino density. In this way, the equation of motion for neutrinos becomes
\begin{equation}
(p^{2}-2\sqrt{2}G_{F}n_{\nu}|\overrightarrow{p}|-M^{2})F_{L}^{-}%
(p)=0,\label{propaequa1}%
\end{equation}
The corresponding equation for antineutrinos (for $\overline{F}^{+}(p)$ ) can
be obtained by replacing the $-$ sign into a $+$ sign in the second term of
this equation.

In order to find the \emph{effective masses,} we have to diagonalize the
matrix $\widehat{M}^{2}\equiv M^{2}\pm2\sqrt{2}G_{F}n_{\nu}|\overrightarrow
{p}|.$ Obviously, this is made by means of \emph{the same rotation angle as in
vacuum}. Hence, the effective masses are
\begin{align}
M_{1}^{2} &  =m_{1}^{2}\pm2\sqrt{2}G_{F}n_{\nu}|\overrightarrow{p}%
|,\nonumber\\
M_{2}^{2} &  =m_{2}^{2}\pm2\sqrt{2}G_{F}n_{\nu}|\overrightarrow{p}%
|,\label{efectmass}%
\end{align}
for each generation, where $m_{1}$ and $m_{2}$ are the vacuum masses, and the
signs $\pm$ correspond, as above, to neutrinos or antineutrinos, respectively.

Let us now assume that, in addition to the self-interaction among the
neutrinos, we have an electrically neutral \emph{background of} \emph{matter}
composed by electrons and nucleons. In this case the effective masses of
neutrinos are \cite{10,11,12}%
\begin{equation}
M_{1,2}^{2}=1/2(A_{c}+\Sigma)\mp1/2(A_{c}^{2}+\Delta^{2})^{1/2}+A_{n},
\end{equation}
where $\ \ \ $%
\begin{align}
\Sigma &  =m_{1}^{2}+m_{2}^{2}\nonumber\\
\ \Delta &  =m_{1}^{2}-m_{2}^{2}\nonumber\\
\ A_{c}  &  =2\sqrt{2}G_{F}|\overrightarrow{p}|n_{e}\nonumber\\
A_{n}  &  =2\sqrt{2}G_{F}|\overrightarrow{p}|n_{\nu}-\sqrt{2}G_{F}%
|\overrightarrow{p}|n_{n} \label{ecupoten}%
\end{align}
$\ \ \ \ \ \ \ \ \ \ \ \ \ \ \ \ \ \ \ \ \ \ \ \ \ \ \ \ \ \ \ \ \ \ \ \ \ \ \ \ \ \ \ \ \ \ \ \ \ \ \ \ \ \ \ \ \ \ \ \ \ \ \ \ \ \ \ \ \ \ \ \ \ \ \ \ \ \ \ \ $%
\ \ \ \ \ \ \ \ \ \ \ \ \ $\ \ \ \ \ \ \ \ \ \ \ \ \ \ \ \ \ \ \ \ \ \ \ \ \ \ \ \ \ \ \ \ \ \ \ \ \ \ \ \ \ \ \ \ \ \ \ \ \ \ \ \ \ \ \ \ \ \ \ $%
\ \ \ \ \ \ \ \ \ \ \ \ \ \ \ \ \ \ \ \ \ \ \ \ \ \ \ \ \ \ \ \ \ \ \ \ \ \ being
$n_{e}$ the number density of electrons (minus antielectrons) and $n_{n}$ the
number density of neutrons (minus antineutrons).

\section{Correction to the Hartree approximation.}

We now want to take into account the effect of the statistical correlations in
our treatment. For this purpose, we define the \emph{two-body correlation
function }for the neutrino fields\emph{\ }as
\begin{equation}
D_{ijkl}^{abcd}(x,x^{\prime},p,p^{\prime})\equiv<\widehat{F}_{ij}%
^{ab}(x,p)\widehat{F}_{kl}^{cd}(x^{\prime},p^{\prime})>-F_{ij}^{ab}%
(x,p)F_{kl}^{cd}(x^{\prime},p^{\prime}),\label{correlations}%
\end{equation}
where subscripts correspond to spin indices. By inserting this definition in
the general equation of motion Eq. (\ref{eqforWig}), and after some
manipulations, it can be written in the form
\begin{align}
\text{ \ \ \ \ } &  \text{\ }[\gamma(\partial-2ip)+2iM]F(x,p)=-(2\pi
)^{-4}\frac{iG_{F}}{2\sqrt{2}}\int d^{4}y^{\prime}d^{4}kd^{4}k^{\prime
}e^{-ik(x-y^{\prime})}\nonumber\\
&  [\Omega Tr\left(  D(x,y^{\prime},p-k/2,k^{\prime})\Psi\right)  +\Omega
F(x,p-k/2)Tr\left(  \Psi F(y^{\prime},k^{\prime})\right)  \nonumber\\
&  +\Lambda Tr\left(  \Phi D(y^{\prime},x,k^{\prime},p-k/2)\right)  +Tr\left(
\Phi F(y^{\prime},k^{\prime})\right)  \Lambda F(x,p-k/2)\nonumber\\
&  +\Omega\left(  Tr\Omega D(y^{\prime},x,k^{\prime},p-k/2)\right)  +Tr\left(
\Omega F(y^{\prime},k^{\prime})\right)  \Omega F(x,p-k/2)\nonumber\\
&  +\Lambda Tr\left(  D(x,y^{\prime},p-k/2,k^{\prime})\Lambda\right)  +\Lambda
F(x,p-k/2)Tr\left(  \Lambda F(y^{\prime},k^{\prime})\right)
].\label{ecuacorrela}%
\end{align}
In Eq. (\ref{ecuacorrela}), the symbol $Tr$ implies summation over both spin
and flavor indexes.

As in the previous section,  we will assume a situation of global equilibrium,
where spatial-time invariance is satisfied. For the correlation functions this
implies that%
\[
D(x,x^{\prime},p,p^{\prime})=D(x-x^{\prime},p,p^{\prime})
\]

When this is applied to the above equation we obtain:
\begin{align}
&  (\gamma p-M)F(p)=\frac{G_{F}}{2\sqrt{2}}\{\gamma^{\mu}(1-\gamma
^{5})F(p)\int d^{4}kTr\left(  \gamma_{\mu}(1-\gamma^{5})F(k)\right)
\nonumber\\
&  +1/2\int d^{4}k^{\prime}d^{4}k[\Omega Tr\left(  \tilde{D}(k,p-k/2,k^{\prime
})\Psi\right)  +\Lambda Tr\left(  \tilde{D}(k,p-k/2,k^{\prime})\Lambda\right)
\nonumber\\
&  +\Lambda Tr\left(  \Phi\tilde{D}(k,k^{\prime},p+k/2)\right)  +\Omega
Tr\left(  \Omega\tilde{D}(k,k^{\prime},p+k/2)]\right)
\},\label{ecuacorrelaequi}%
\end{align}
where the function $\tilde{D}(k,p,p^{\prime})$ is the Fourier transformed of
the correlation function
\begin{equation}
\tilde{D}(k,p,p^{\prime})=\frac{1}{(2\pi)^{-4}}\int d^{4}xe^{-ikx}%
D(x,p,p^{\prime}).\label{transcorrela}%
\end{equation}

We will impose the same three hypothesis as in the Hartree approximation,
which implies having definite mass eigenstates and a mixing angle, which mixes
the mass eigenstates to produce the flavor eigenstates of neutrinos. Moreover,
under these conditions, we can apply the Wick's theorem (see, for example
\cite{FW}) to calculate correlation functions in terms of the Wigner Function
of the neutrino fields, as derived in the appendix. We obtain:
\begin{equation}
\tilde{D}_{ijkl}^{abcd}(k,p,p^{\prime})=-\delta^{4}(p-p^{\prime})F_{Lkj}%
^{cb}(p^{\prime}+k/2)F_{Lil}^{ad}(p^{\prime}-k/2), \label{calcucorrela}%
\end{equation}
where we are only using the left projections of the neutrino fields in order
to construct the Wigner Functions, and hence the correlation functions. By
inserting the above relation in the equation of motion, this one is finally
left as
\begin{equation}
\lbrack\gamma p-\sqrt{2}G_{F}\gamma a+\sqrt{2}G_{F}\int d^{4}q\gamma
F_{L}(q)\gamma-M]F_{L}(p)=0. \label{ecuacorrelaequi1}%
\end{equation}
The four-vector $a$ is defined in the same way as in the previous section.
Obviously, the third term of this equation provides us with an additional
correction to the corresponding equation in the Hartree approximation.

\section{Propagation modes of the neutrinos.}

Starting from the latter equation, we can now calculate the propagation modes
of the neutrinos in the gas, by performing the following decomposition,
consistent with the hypothesis made in the previous sections
\begin{equation}
F_{L}(q)=F_{L}^{-}(q)+\bar{F}_{L}^{+}(q)=U^{\ast}(q)F_{L}^{-\ast}(q)U^{^{\ast
}\dagger}(q)+\bar{U}^{\ast}(q)\bar{F}_{L}^{+\ast}(q)\bar{U}^{\ast\dagger
}(q),\label{WignerFunc}%
\end{equation}
that is, both the corresponding part of neutrinos $F_{L}^{-}(q)$ and the
corresponding part of antineutrinos $\bar{F}_{L}^{+}(q)$ of the Wigner
Function with flavor indices can be expressed in terms of Wigner functions
with effective mass indices $F_{L}^{-\ast}(q)$ and $\bar{F}_{L}^{+\ast}(q)$
for neutrinos and antineutrinos, respectively (whose mass eigenstates will not
be the same as in the Hartree approximation), by means of the unitary
transformation $U^{\ast}(q)$ or $\bar{U}^{\ast}(q)$, each of them defined by
the corresponding rotation angle.

On the other hand, we define the following quantities:
\begin{align}
&  n_{\nu_{e}}=\frac{1}{2\pi^{2}}\int_{0}^{\infty}d|\vec{q}||\vec{q}%
|^{2}[c^{2}f_{1}(q)+s^{2}f_{2}(q)],\nonumber\\
&  n_{\nu\mu}=\frac{1}{2\pi^{2}}\int_{0}^{\infty}d|\vec{q}||\vec{q}|^{2}%
[s^{2}f_{1}(q)+c^{2}f_{2}(q)],\nonumber\\
&  n_{\nu_{12}}=\frac{1}{2\pi^{2}}\int_{0}^{\infty}d|\vec{q}||\vec{q}%
|^{2}cs[f_{1}(q)-f_{2}(q)], \label{numdens}%
\end{align}
where $f_{1}(q)$ and $f_{2}(q)$ are the \emph{Fermi statistical distribution
functions} for each generation, corresponding to quasi-particles with
well-defined effective masses, and $s$ and $c$ are the $sin$ and $cos$ of the
rotation angle $\theta$, which relates the eigenstates of effective masses to
flavor eigenstates. In this way, $n_{\nu_{e}}$ is the number density of
electron neutrinos, $n_{\nu_{\mu}}$is the number density of muon neutrinos and
$n_{\nu_{12}}$ contains interference effects. Analogously, we can define the
number densities for antineutrinos $\bar{n}_{\nu_{e}}$, $\bar{n}_{\nu_{\mu}}$
and $\bar{n}_{\nu_{12}}$.

Carrying out a straightforward calculation, we arrive to
\begin{equation}
\int d^{4}qF_{L}(q)=\frac{1}{2}NP_{L}\gamma^{0}, \label{calcuWigner}%
\end{equation}
where $P_{L}$ is the left chirality projector and $N$ is a matrix defined (in
flavor space) as
\begin{equation}
N=\left(
\begin{array}
[c]{cc}%
n(e) & n_{12}\\
n_{12} & n(\mu)
\end{array}
\right)  =\frac{1}{2}\left(
\begin{array}
[c]{cc}%
n_{\nu} & 0\\
0 & n_{\nu}%
\end{array}
\right)  +\frac{1}{2}\left(
\begin{array}
[c]{cc}%
\delta & 2n_{12}\\
2n_{12} & -\delta
\end{array}
\right)  , \label{matrixN}%
\end{equation}
in which $n(e)=n_{\nu_{e}}-\bar{n}_{\nu_{e}}$ is the electron neutrino (minus
electron antineutrino) number density, $n(\mu)=n_{\nu_{\mu}}-$ $\bar{n}%
_{\nu_{\mu}}$ is the muon neutrino (minus muon antineutrino) number density,
$n_{\nu}=n(e)+n(\mu)$ is the total number neutrino density, $n_{12}%
=n_{\nu_{12}}-$ $\bar{n}_{\nu_{12}}$, and $\delta=n(e)-n(\mu)$ is a
statistical parameter of asymmetry between the two flavors. It is important to
note that the above expressions exactly coincide with those obtained in
\cite{3p} by using a totally different method. We can finally write the
equation of motion in the form
\begin{equation}
(\gamma p-\sqrt{2}G_{F}\gamma^{0}n_{\nu}-\sqrt{2}G_{F}N\gamma^{0}%
-M)F_{L}(p)=0. \label{ecuacorrela2}%
\end{equation}
Obviously, if $\delta\neq0$ and $n_{12}\neq0$ then the interacting mixing
angle $\theta$ will not be the same as the vacuum mixing angle $\theta_{0}$.
At this point, we can add the contribution from an \emph{electrically neutral
background} of protons, neutrons and electrons. This amounts to replacing the
matrix $N$ by
\begin{equation}
N^{\ast}=\frac{1}{2}\left(
\begin{array}
[c]{cc}%
n_{\nu}+\delta+2n_{e}-n_{n} & 2n_{12}\\
2n_{12} & n_{\nu}-\delta-n_{n}%
\end{array}
\right)
\end{equation}

In this way, Eq. (\ref{ecuacorrela2}) will now become:%

\begin{equation}
(\gamma p-\sqrt{2}G_{F}\gamma^{0}n_{\nu}-2\sqrt{2}G_{F}N^{\ast}\gamma
^{0}-M)F_{L}(p)=0. \label{ecuacorrela4}%
\end{equation}
To find the propagation modes, we multiply Eq. (\ref{ecuacorrela4}) by
$(\gamma p-\sqrt{2}G_{F}\gamma^{0}n_{\nu}-2\sqrt{2}G_{F}N^{\ast}\gamma^{0}+M)$
, and neglect the terms of order $G_{F}^{2}$, as in the Hartree approximation.
Then, one obtains:
\begin{equation}
(p^{2}-\sqrt{2}G_{F}n_{\nu}p^{0}-\sqrt{2}G_{F}N^{\ast}p^{0}-M^{2})F_{L}(p)=0
\label{ecuacorrela3}%
\end{equation}

In order to find the neutrino effective masses we have to diagonalize the
matrix
\begin{equation}
\hat{M}^{2}=M^{2}+\sqrt{2}G_{F}n_{\nu}p^{0}+2\sqrt{2}G_{F}N^{\ast}p^{0},
\label{massmatrix}%
\end{equation}
and to find the mixing angle in matter we have to obtain the unitary
transformation, which is determined by the corresponding rotation. After some
algebra, we arrive to the following expression for the effective masses in the
medium:
\begin{equation}
M_{1,2}^{\ast2}\left(  p^{0}\right)  =\frac{1}{2}\left[  \Sigma+2\sqrt{2}%
G_{F}\left(  3n_{\nu}+n_{e}-n_{n}\right)  p^{0}\right]  \mp\frac{1}{2}%
\Delta^{\ast}, \label{efectivemass}%
\end{equation}
where
\begin{equation}
\Delta^{\ast}=\left[  \left(  2\sqrt{2}G_{F}p^{0}\left(  n_{e}+\delta\right)
-\Delta\cos2\theta_{0}\right)  ^{2}+\left(  4\sqrt{2}G_{F}p^{0}n_{12}%
-\Delta\sin2\theta_{0}\right)  ^{2}\right]  ^{1/2}%
\end{equation}
is the effective mass difference. In Eq. (\ref{efectivemass}) the upper
(lower) sign corresponds to $M_{1}^{\ast}$ ($M_{2}^{\ast}$) ,where $\Sigma$
and $\Delta$ have been defined in Eq. (\ref{ecupoten}). The mixing angle is
given by
\begin{equation}
\sin2\theta=\frac{\Delta\sin2\theta_{0}-4\sqrt{2}G_{F}p^{0}n_{12}}%
{\Delta^{\ast}}. \label{anglematter}%
\end{equation}
Notice that Eqs. (\ref{efectivemass}-\ref{anglematter}) depend on $n_{\nu}$
and $\delta$ which, in turn, have to be evaluated using the above equations.
In other words, the whole set of equations has to be solved
\emph{self-consistently}.

Eq. (\ref{ecuacorrela3}) gives the dispersion relation for neutrinos and
antineutrino mass eigenstates, which can be written as:
\begin{equation}
p^{2}-M_{1,2}^{\ast2}\left(  p^{0}\right)  =0
\end{equation}
and provides (as an implicit equation) the energy $p^{0}$ as a function of the
momentum $|\vec{p}|$. As a first approximation, one can use the fact that,
under most situations of interest, neutrinos are extremely relativistic
particles. Thus, for neutrinos one can replace $p^{0}$ by $|\vec{p}|$. In this
way, the above dispersion equation can be approximately solved as:
\begin{equation}
p_{0}=\sqrt{|\vec{p}|^{2}+M_{1,2}^{\ast2}\left(  |\vec{p}|\right)  }%
\end{equation}
To obtain the corresponding formulae for antineutrinos we only have to change
$|\vec{p}|$ to $-|\vec{p}|$ in the previous equation.

A consequence of Eq. (\ref{anglematter}) is that the condition for the MSW
resonance is modified with respect to the situation where there is not a
neutrino background. In fact, the condition for the resonance is now:
\begin{equation}
p^{0}=\frac{\Delta\left[  \left(  \delta+n_{e}\right)  \cos2\theta_{0}%
+2n_{12}\sin2\theta_{0}\right]  }{2\sqrt{2}G_{F}\left[  \left(  \delta
+n_{e}\right)  ^{2}+4n_{12}^{2}\right]  } \label{MSWmod}%
\end{equation}

This new condition can be of interest if $\sin2\theta_{0}\simeq1$, as
suggested for both solar and atmospheric neutrino oscillation values. In this
case, the MSW resonance might be dominated by the neutrino background. In
order to investigate this possibility, we need to calculate $n_{12}$ and
$\delta$. Using Eqs. (\ref{numdens}) and (\ref{anglematter}) we arrive to the
formulae%
\begin{align}
n_{12}  &  =\frac{\Delta\sin(2\theta_{0})F_{1}}{1+4\sqrt{2}G_{F}F_{2}%
}\nonumber\\
\delta+n_{e}  &  =\frac{n_{e}+2\Delta\cos(2\theta_{0})F_{1}}{1+4\sqrt{2}%
G_{F}F_{2}} \label{dandn12}%
\end{align}

where%
\begin{align}
F_{1}  &  =\frac{1}{4\pi^{2}}\int_{0}^{\infty}dpp^{2}\frac{f_{1}(p)-f_{2}%
(p)}{\Delta^{\ast}(p)}\nonumber\\
F_{2}  &  =\frac{1}{4\pi^{2}}\int_{0}^{\infty}dpp^{3}\frac{f_{1}(p)-f_{2}%
(p)}{\Delta^{\ast}(p)} \label{F1F2}%
\end{align}

The numerator in the integrand of the above equations is a small quantity, due
to the small mass difference $\Delta^{\ast}(p)$, therefore it is convenient to
expand the numerator using $\Delta^{\ast}(p)$ as a parameter. We then obtain:%
\[
f_{1}(p)-f_{2}(p)\simeq-\frac{\Delta^{\ast}(p)}{2p}\frac{\partial f_{1}%
(p)}{\partial p}%
\]

Moreover, in the extremely relativistic limit, we can write
\[
f_{1}(p)\simeq\frac{1}{1+\exp\left[  \left(  p-\mu\right)  /T\right]  }%
\]

In this way, the integrals in Eq. (\ref{F1F2}) can be performed, giving%
\begin{align}
F_{1}  &  =\frac{T}{8\pi^{2}}\ln\left[  1+\exp\left(  \mu/T\right)  \right]
\nonumber\\
F_{1}  &  =-\frac{T^{2}}{4\pi^{2}}L_{i2}\left[  -\exp\left(  \mu/T\right)
\right]
\end{align}

with $L_{i2}$ the dilogarithmic function.

The correction to the MSW condition, as given in Eq. (\ref{MSWmod}), is then
of the order%
\[
\frac{n_{12}}{n_{e}}\tan2\theta_{0}%
\]
Let us consider the conditions in the Early Universe, just before the
nucleosynthesis. We then have a temperature $T\sim1$ MeV , $\mu\sim0.1$ MeV
and $n_{e}\sim0.1MeV^{3}$. By substituting into the above expression, one
finds that the correction to the MSW condition is only meaningful if%
\[
\tan2\theta_{0}>10^{15}%
\]
which implies $\theta_{0}=\pi/4$. Such values seem to be disfavored (although
not excluded) for $\nu_{e}\rightarrow(\nu_{\mu},\nu_{\tau})$ oscillations
\cite{14}. Only if this value is allowed, the neutrino background can play a
role in establishing the MSW condition. On the other hand, one can check that
the corrections of the neutrino background to the effective masses, as given
by Eq. (\ref{efectivemass}), are negligible.

Another possibility consists in the adiabatic $\nu_{\mu}\rightarrow\nu_{\tau}$
conversion in the presence of a neutrino background. The corresponding
resonance formula can be found to be
\begin{equation}
p^{0}=\frac{\Delta\left[  \delta\cos2\theta_{0}+2n_{12}\sin2\theta_{0}\right]
}{2\sqrt{2}G_{F}\left[  \delta^{2}+4n_{12}^{2}\right]  }%
\end{equation}
As an example, we consider the proto-neutron star interior with $T\sim40$ MeV
and $\mu\sim100$ MeV. We then find $p^{0}\sim10^{9}$ MeV which is, of course,
too high for the neutrinos produced in such a scenario.

\section{Conclusions.}

In this paper, we have investigated the equilibrium properties of a system of
two generations of mixed massive Dirac neutrinos in equilibrium, when
self-interactions are taken into account. To this end, we have used techniques
based on Wigner functions. We assume that well-defined quasi-particle states
exist for the neutrinos, i.e. the fermion fields for states with a definite
mass can be expanded in plane-wave states, similarly to the non-interacting
case, but with a different dispersion relation.

The equilibrium state is characterized by a single chemical potential $\mu$
and a temperature $T$. First, we analyzed the Hartree approximation (when
correlations are neglected). In this case, self-interactions are diagonal in
flavor space and do not modify the mixing angle, although they change the
effective masses.

The inclusion of correlations can be done, under the conditions assumed above,
using the derivations made in appendix. Our results for these corrections
agree with previous calculations \cite{3s,3p}, using completely different
techniques. These corrections give a \emph{non-diagonal} term in the effective
mass matrix. Therefore, in addition to a modification in the effective masses
of eigenstates, there is a change in the in-medium mixing angle, as compared
to the Hartree result. Also, the condition for the MSW resonance (when a
matter background is included) differs from the usual MSW condition. We have
shown, however, that, for typical astrophysical and cosmological scenarios,
the corrections are small, although they can be of some interest in cosmology
if the $\nu_{e}$-($\nu_{\mu},\nu_{\tau}$) mixing angle is exactly $\pi/4$, as
still allowed by neutrino oscillation experiments and theoretical models
\cite{14,15}.

\begin{acknowledgments}
This work has been supported by the Spanish grants FPA2002-00612 \ and
AYA2001-3490-C02. We are indebted to Sergio Pastor for a fruitful discussion
and comments.
\end{acknowledgments}

\bigskip\appendix

\section{Calculation of correlations.}

In this appendix we calculate the statistical correlations defined by Eq.
(\ref{correlations}) under the quasi-particle hypothesis. Let us consider, for
simplicity, that we have a neutrino field in equilibrium consisting on one
generation of massive neutrinos. Using the corresponding unitary
transformation to mass eigenstates, the following procedure can be easily
generalized to more than one generation. The Wigner function for this field
is
\begin{equation}
\widehat{F}_{ij}(x,p)=(2\pi)^{-4}\int d^{4}y\,e^{-ipy}<\widehat{\overline
{\Psi}}_{j}(x+y/2)\widehat{\Psi}_{i}(x-y/2)>,
\end{equation}
and the \emph{correlation functions} can be expressed as
\begin{equation}
D_{ijkl}(x,x^{\prime},p,p^{\prime})=<\widehat{F}_{ij}(x,p)\widehat{F}%
_{kl}(x^{\prime},p^{\prime})>-F_{ij}(x,p)F_{kl}(x^{\prime},p^{\prime})
\label{defcorrapen}%
\end{equation}
We can now evaluate the first term on the right hand using Wick's theorem
\cite{FW}. In this way, we arrive to:
\begin{align}
&  <\widehat{F}_{ij}(x,p)\widehat{F}_{kl}(x^{\prime},p^{\prime})>=(2\pi
)^{-8}\int d^{4}yd^{4}y^{\prime}\,e^{-ipy}e^{-ip^{\prime}y^{\prime}}\\
&  <\widehat{\overline{\Psi}}_{j}(x+y/2)\widehat{\Psi}_{i}(x-y/2)\widehat
{\overline{\Psi}}_{l}(x^{\prime}+y^{\prime}/2)\widehat{\Psi}_{k}(x^{\prime
}-y^{\prime}/2)>\nonumber\\
=  &  (2\pi)^{-8}\int d^{4}yd^{4}y^{\prime}\,e^{-ipy}e^{-ip^{\prime}y^{\prime
}}[\nonumber\\
&  <\widehat{\overline{\Psi}}_{j}(x+y/2)\widehat{\Psi}_{i}(x-y/2)><\widehat
{\overline{\Psi}}_{l}(x^{\prime}+y^{\prime}/2)\widehat{\Psi}_{k}(x^{\prime
}-y^{\prime}/2)\nonumber\\
-  &  <\widehat{\overline{\Psi}}_{j}(x+y/2)\widehat{\overline{\Psi}}%
_{l}(x^{\prime}+y^{\prime}/2)><\widehat{\Psi}_{i}(x-y/2)\widehat{\Psi}%
_{k}(x^{\prime}-y^{\prime}/2)\nonumber\\
+  &  <\widehat{\overline{\Psi}}_{j}(x+y/2)\widehat{\Psi}_{k}(x^{\prime
}-y^{\prime}/2)><\widehat{\Psi}_{i}(x-y/2)\widehat{\overline{\Psi}}%
_{l}(x^{\prime}+y^{\prime}/2)>].
\end{align}
The first term in this expression leads to a product of Wigner functions which
is canceled by the second term in Eq. (\ref{defcorrapen}). The second term
vanishes. Finally, the third term can be expressed as:
\begin{align}
(2\pi)^{-8}  &  \int d^{4}yd^{4}y^{\prime}\,e^{-ipy}e^{-ip^{\prime}y^{\prime}%
}<\widehat{\overline{\Psi}}_{j}(x+y/2)\widehat{\Psi}_{k}(x^{\prime}-y^{\prime
}/2)>\nonumber\\
\times &  \left[  -<\widehat{\overline{\Psi}}_{l}(x^{\prime}+y^{\prime
}/2)\widehat{\Psi}_{i}(x-y/2)>-iS_{il}(x-y/2-x^{\prime}-y^{\prime}/2)\right]
\end{align}
In this formula $S_{il}()$ is the propagator of the neutrino field. It appears
because normal ordering of the operators has no been considered. We now impose
normal ordering, which amounts to neglecting the last term. In this way, we
find:
\begin{align}
&  D_{ijkl}(x,x^{\prime},p,p^{\prime})=-(2\pi)^{-8}\int d^{4}yd^{4}y^{\prime
}\,e^{-ipy}e^{-ip^{\prime}y^{\prime}}\nonumber\\
&  <\widehat{\overline{\Psi}}_{j}(x+y/2)\widehat{\Psi}_{k}(x^{\prime
}-y^{\prime}/2)><\widehat{\overline{\Psi}}_{l}(x^{\prime}+y^{\prime
}/2)\widehat{\Psi}_{i}(x-y/2)>
\end{align}

By using the equality
\begin{equation}
\widehat{\overline{\Psi}}_{i}(x+z/2)\widehat{\Psi}_{j}(x-z/2)=\int
d^{4}p\,e^{ipz}\widehat{F}_{ji}(x,p),
\end{equation}
and the conditions of equilibrium, which imply that $F(x,p)=F(p),$ we obtain
\begin{equation}
D_{ijkl}(x-x^{\prime},p,p^{\prime})=-\int d^{4}q\delta(p-p^{\prime
})e^{-i2(p^{\prime}-q)(x-x^{\prime})}F_{kj}(q)F_{il}(2p^{\prime}-q).
\label{29}%
\end{equation}
If we take the Fourier transformation of the correlations, we have that
\begin{equation}
\widetilde{D}_{ijkl}(k,p,p^{\prime})=(2\pi)^{-4}\int d^{4}(x-x^{\prime
})D_{ijkl}(x-x^{\prime},p,p^{\prime})e^{-ik(x-x^{\prime})}%
\end{equation}
and we end up with the following result:
\begin{equation}
\widetilde{D}_{ijkl}(k,p,p^{\prime})=-\delta(p-p^{\prime})F_{kj}(p^{\prime
}+k/2)F_{il}(p^{\prime}-k/2). \label{31}%
\end{equation}

\end{document}